\documentclass{iopart}

\usepackage{epsfig}

\eqnobysec

\begin{document}

\jl{1}

\title[Orientational order on curved surfaces]{Orientational order
on curved surfaces - the high temperature region}

\author{Georg Foltin\dag\  and Raphael A Lehrer\ddag}
\address{\dag\ Institut f\"ur Theor\-etische
Physik IV, Hein\-rich\---Heine--Uni\-ver\-sit\"at D\"us\-sel\-dorf,
Uni\-ver\-sit\"ats\-strasse
1, D--40225 D\"ussel\-dorf, Germany}

\address{\ddag\ Lyman Laboratory of Physics, Harvard University,
Cambridge, Massachusetts 02138}

\begin{abstract}
We study orientational order, subject to thermal fluctuations, on a
fixed curved surface. We derive, in particular, the average density of
zeros of Gaussian distributed vector fields on a closed Riemannian
manifold.  Results are compared with the density of disclination
charges obtained from a Coulomb gas model. Our model describes the
disordered state of two dimensional objects with orientational degrees
of freedom, such as vector ordering in Langmuir monolayers and
lipid bilayers above the hexatic to fluid transition.
\end{abstract}


\section{Introduction}
\label{sec-intro}

In several areas of statistical physics and condensed matter, a great
deal of progress has been achieved by focusing on the physics of
topological defects, ignoring other degrees of freedom.  The
Kosterlitz-Thouless transition, describing the destruction of
orientational order in thin films, is a particularly important
example.  The transition is viewed as one where defect pairs unbind
and proliferate, destroying the (quasi)-long-range order~\cite{KT}.
Another crucial example is that of type-II superconductors in a
magnetic field, where the important physics is encoded in the
properties of vortex lines~\cite{Blatter}. Even in the absence of a
magnetic field, the formation and growth of vortex loops can be used
to explain the form of the voltage vs.\ current relation~\cite{HFF}.
Yet a third example is found in the physics of orientational order in
membranes, where many workers~\cite{OT,PLM,Evans,park1} have found it
fruitful to focus on the properties of topological defects to
understand the low-temperature physics.

In all of these examples, topological defects are used primarily to
understand the low-temperature behavior.  For example, in the case of
superconductors, the vortex line description is used primarily to
understand the behavior below $T_\mathrm{c}$ (or $H_{\mathrm{c2}}$),
rather than to
understand the properties of the normal metal phase at higher
temperatures.  Indeed, topological defects are a more natural
description at low temperatures, where it is very costly to excite the
order parameter away from its average value.  Although themselves
energetically costly, topological defects are the minimal-energy way
of satisfying a constraint of the system - e.g., the
curvature of a membrane or the penetration of a magnetic field into a
superconductor.

At higher temperatures, the description becomes less natural.  Order
parameter fluctuations become much less costly, and hence the
fluctuations observed in thermal equilibrium become more violent.  At
sufficiently high temperatures, the broken symmetry associated with
the low-temperature phase is restored, and the average value of the
order parameter is zero.  Above this temperature, the order parameter
behaves approximately like a Gaussian random variable.  In this case,
it seems that the description in terms of topological defects will be
insufficient to describe the physics, as it only encodes the locations
(and signs) of the zeros of the order parameter, classifying as
irrelevant any fluctuations that the order parameter undergoes between
these zeros.

Despite these objections, a description in terms of topological
defects actually approximates the high temperature behavior of such
systems quite well.  For the case of thin films,
Halperin~\cite{halperin} showed that the density of topological
defects (i.e. zeros of the order parameter) that one obtains
from a Gaussian order parameter are much the same as one obtains from
a Coulomb gas model that allows only the topological defects as
degrees of freedom.  In the case of superconductors, Lehrer and
Nelson~\cite{LN} have shown that above $H_{\mathrm{c2}}$, a Gaussian
approximation to the Ginzburg-Landau free energy predicts
approximately the same distribution of vortex loops and lines as does
the London theory, which is purely a description in terms of
topological defects.

In this paper, we focus on the case of membranes.  In particular, we
examine the density of topological defects under the approximation
that the free energy is Gaussian.  This approximation will be valid at
high temperatures.  We compare these results to results obtained from
a model which focuses \textit{only} on the topological defects and
their interactions, namely, a Debye-H\"uckel approximation to a
Coulomb gas model. 

One way to probe the high-temperature properties of a topologically spherical
surface is with light scattering experiments on lipid vesicles
\cite{giant}.
In the case of lipid bilayers, the source of the orientational order
parameter is the vector between a lipid head and a neighboring head, and the
order parameter describes the tendency of the lipids to achieve
hexatic order at low temperatures.
(The hexatic order and the  transition to the disordered phase were
studied in free standing liquid-crystal
films using light and X-ray scattering \cite{ninehalf}.)
However, since this order parameter
is invariant under rotations by $60^0$ (because most lipids have six
neighbors on average),
it is slightly different from the case we consider, where
the order parameter is only invariant under rotation by integer
multiples of $360^0$. Nevertheless, we expect much of what
we derive here to apply to these systems.
 
\begin{figure}
\begin{center}
\epsfig{file=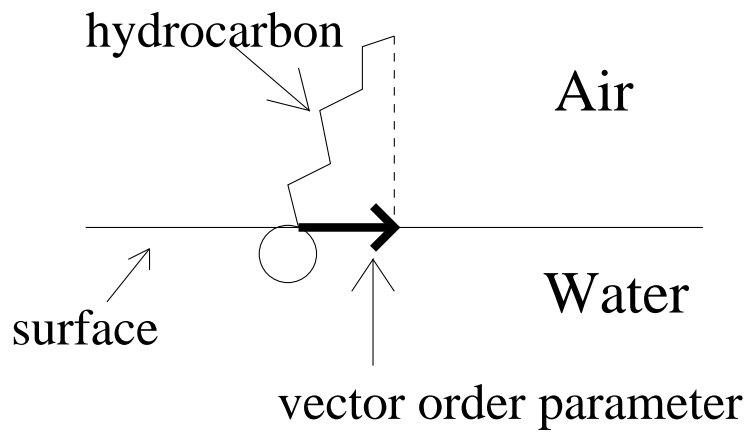,width=10.cm}
\end{center}
\caption{\label{fig2}A surfactant molecule tilted away from the normal.}
\end{figure}

A system that is closer to what we consider
here is that of tilted Langmuir monolayers \cite{langmuir},
which consist of a monolayer of lipids or amphiphiles  on a liquid
surface, {\it e.g.} the surface of a water droplet.
The surfactant molecules have a tendency to orient themselves so that
the hydrocarbon chain is at a tilt away from the normal to the surface.
The projection of the direction of the polymer chain onto
the surface forms an orientational order parameter which is exactly of the form
that we consider in this paper, as illustrated in \fref{fig2}. A
similar situation may be obtained in a lipid bilayer when the lipids tend
to tilt away from the normal to the surface \cite{gennes,chiang}, providing
yet another source for an orientational order parameter. 

The rest of the paper is organized as follows.  In
\sref{sec-coulomb}, we write down a continuum model for the orientational
order parameter that has the correct symmetries and should describe
both the low- and the high-temperature physics of the orientational
order parameter, namely, a $O(2)-$Ginzburg-Landau theory in curved space.
From this, we will derive a Coulomb gas model, and use it to calculate the
density of disclination defects in the Debye-H\"uckel approximation in
the high-temperature limit.  We obtain
\begin{equation}
2\pi\rho=K+\frac{1}{4\pi^2 K_{\mathrm{A}} x}
\Delta K+\Or (x^{-2}),
\end{equation}
where $\rho$ is the density of defects, $K$ is the curvature of the
surface, $\Delta$ is the Laplace-Beltrami operator,
$x$ is the fugacity of the topological defects, and $K_{\mathrm{A}}$
describes the interaction strength of the defects. 

In \sref{sec-gauss}, we present the main results of our paper. We approximate 
our model by neglecting the nonlinear terms, valid at high
temperatures, and calculate the density of defects for an arbitrarily curved
surface.
For simplicity, we restrict our scope to closed
surfaces - namely surfaces that are topologically
equivalent to spheres, tori, etc.  We especially focus on the case
where the surface is topologically equivalent to a sphere, both for
calculational ease and because we expect this class of (closed) surfaces to be
the most easily amenable to experiments.
Because the calculation is fairly technical, we first review Halperin's
calculation of the density of topological defects that we expect to
see in flat space as a ``toy model'' for the problem in curved space.
We then proceed to the calculation in curved space, obtaining
\begin{equation}
2\pi\rho=
K+\frac{\Delta K}{12\pi Z\tau}+\Or (\tau^{-2}),
\end{equation}
where $\tau$ measures the deviation from the critical temperature,
$Z^{-1}=2\pi/\log(1/(a^2\tau))$, and $a$ is a short distance
cutoff. Instead of using a gauge-field
representation of the orientational order parameter we deal with a manifestly 
gauge-invariant picture. Employing
a special symmetry of the model we can express the director field through
simple scalar fields and solve it exactly in a high-temperature expansion.

The result for the defect density in curved space is
equivalent to the Coulomb gas/Debye-H\"uckel result
above, provided we identify $\pi K_{\mathrm{A}}x=3Z\tau$. 
This confirms the validity
of the Coulomb model even for high temperatures at this level of
approximation. Deviations do, however, begin to show up at $\Or (\tau^{-2})$,
as we shall show below.

\section{Model and Debye-H\"uckel theory}
\label{sec-coulomb}

We concentrate on the case of purely in-plane orientational order and
therefore use a (two component) tangential vector field $u_i(\sigma)$
as the order parameter.  To describe the physics of the surface, we
rely on the language of Riemannian differential geometry, which
ensures that results are independent of any particular coordinate
system.  (A concise introduction to differential geometry of surfaces
can be found in reference~\cite{nels}.)  The order parameter lives on a
closed two-dimensional Riemannian manifold with line element
$\rmd s^2=g_{ij}\rmd\sigma^i\rmd\sigma^j$, where
$g_{ij}=g_{ij}(\sigma^1,\sigma^2)$ is the metric tensor and
$\sigma=(\sigma^1,\sigma^2)$ are internal coordinates of the surface.
Using this formalism, we write down a $O(2)-$invariant, statistical
weight $P[u]\propto\exp(-H/T)$ for the $u_i-$field, where $H$ is the
(mesoscopic) Hamiltonian and $T$ is the temperature.
The simplest such Hamiltonian $H$ is the
analogue of Ginzburg-Landau theory in flat space, namely
\begin{equation}
\label{mexican}
\frac{H}{T}=\frac{1}{2}\int\rmd A\left(
D_iu_jD^iu^j+\tau u_iu^i+c(u_iu^i)^2\right),
\end{equation}
where $\rmd A=\sqrt{g}(\sigma)\;\rmd\sigma^1\rmd\sigma^2$
is the invariant area element,
$g$ is the determinant of the metric, $D_i$ is the covariant gradient,
$u^i=g^{ij}u_j$, $g^{ij}=(g^{-1})_{ij}$,
$c$ is the coupling constant, and the coupling
constant for the gradient term is absorbed into the field $u_i$.
\Eref{mexican} encodes the same physics as the free
energy used by Park \textit{et\ al\ }~\cite{PLM} and by
Evans~\cite{Evans} for vector defects, although theirs appears in the
gauge-field picture. The equivalence of the models is shown explicitly
in \ref{gauge}.

For $\tau$ below a certain critical $\tau$, this will be a ``Mexican
hat potential''; however, we will be concerned primarily with the
opposite case of high temperatures ($\tau \gg 0$).  Before we
specialize to this case, we look at some properties at smaller $\tau$.
A critical temperature $\tau_c$ (the mean field value $\tau_c=0$ gets
renormalized due to fluctuations) separates the disordered state
$\tau>\tau_c$ (high-temperature region) from the ``ordered state''
$\tau<\tau_c$.  We put ``ordered state'' in quotation marks, because a
perfectly ordered state is impossible for certain manifold topologies.
For example, on a sphere or any other surface with the same topology,
a tangential vector field has at least two zeros
(defects)~\cite{nels}. This can be illustrated by attempting to comb a
hedgehog or a hairy ball: there will be two places where the vector
field is zero or has a singularity.

\begin{figure}
\begin{center}
\epsfig{file=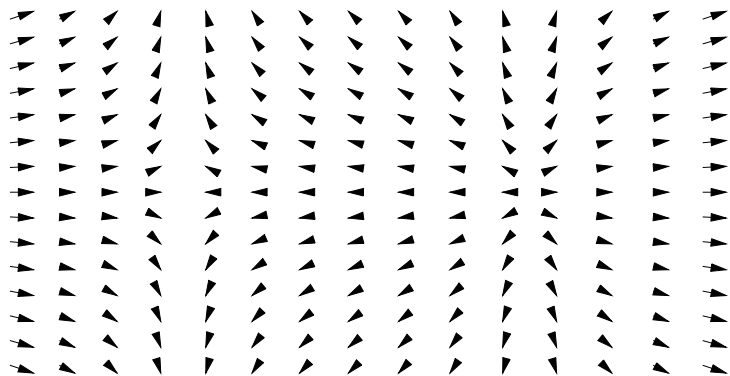,width=10.cm}
\end{center}
\caption{\label{fig1}The planar vector field $(u_x,u_y)=(x^2-1,y)$
with a negative zero (left) and a positive zero (right).}
\end{figure}

To investigate this in more detail, we distinguish between two types
of zeros.  One type, called a ``positive zero'', is characterized by
$\det(D_iu_j)>0$, while the other type, a ``negative zero'', has a
saddle-like flow and is characterized by $\det(D_iu_j)<0$.  See
\fref{fig1} for an illustration of these types of defects.  
Zeros with $\det(D_iu_j)=0$ do not fall into this scheme; however,
they will not show up in a statistical model as the probability to hit
exactly $\det(D_iu_j)=0$ vanishes. 
The number of positive zeros minus the number of negative zeros is a
topological constraint and equal to $2(1-\gamma)$, where $\gamma$ is
the number of handles of the (closed) surface, e.g. zero for
a spherical topology and one for a torus~\cite{nels}. We will show
this theorem explicitly en route to our calculation.  Thus, on
surfaces other than tori, the low-temperature phase necessarily has
defects, unlike in flat space, where the ground state is defect free.

As in flat space, the properties of the low temperature phase are
determined by low energy Goldstone modes (``spin waves''), which
prevent true long ranged correlations.  Instead, one finds an
algebraic decay of the correlations (quasi-long-ranged order).
Besides the spin waves, thermally excited defects persist.
Integration over the spin waves results in a Coulomb gas model for
these defects (zeros)~\cite{park1}, where the defects carry a charge
proportional to their index $q=\mbox{sign}\det(D_iu_j)$ and a core
energy (chemical potential~\cite{deem}).  The interaction energy of
the defects reads 
\begin{equation}
\label{cgm}
\frac{H}{T}=\frac{K_{\mathrm{A}}}{2}\int\rmd A\int\rmd A'\;(2\pi\rho-K)_\sigma
\;G(\sigma,\sigma')\;(2\pi\rho-K)_{\sigma'},
\end{equation}
where $G(\sigma,\sigma')$ denotes the Green's function of the negative
of the Laplace-Beltrami operator $ - \Delta= - g^{ij}D_i\partial_j$
\footnote{
The Greens's function of the negative of the Laplace-Beltrami operator
$\Delta$ is the ``electrostatic''
potential at the point $\sigma$ of a unit charge located at point $\sigma'$
and of a negative unit charge which is uniformly
distributed over the surface to ensure charge neutrality. It is given by
$-\Delta G(\sigma,\sigma')=\delta_{\mathrm{c}}(\sigma,\sigma')-1/A$, where $\delta_{\mathrm{c}}$
is the covariant delta function and $A$ is the area
of the surface. 
}
and $K_{\mathrm{A}}$ is the coarse-grained effective coupling between
the defects and therefore depends on the
temperature~\cite{KT,nels2}. $\rho$ is the defect density
$\rho(\sigma)=\sum_i q_i\;\delta_{\mathrm{c}}(\sigma,\sigma_i)$,
where $\sigma_i$
are the locations of the defects and $\delta_{\mathrm{c}}$
is the covariant version of the Dirac delta function given by
\[\delta_{\mathrm{c}}(\sigma,\sigma')=\lim_{\lambda\rightarrow\infty}
\frac{\lambda}{2\pi}\exp\left(-\frac{\lambda}{2}d^2(\sigma,\sigma')\right)=
\delta^2(\sigma-\sigma')/\sqrt{g}(\sigma)\]
($d(\sigma,\sigma')$ is the geodesic distance between $\sigma$ and
$\sigma'$). We note that the
Gaussian curvature $K=K(\sigma)$ plays the role of a background charge
density.  Because of this, positive defects tend to concentrate in
regions with positive curvature, whereas negative prefer saddle shaped
regions with negative curvature.  Charge neutrality and the
Gauss-Bonnet theorem~\cite{nels} $2\pi \sum_i q_i-\int\rmd A\;K=0$ yield
the topological constraint $\sum_i q_i=2(1-\gamma)$.

Above a certain temperature it is expected~\cite{park1} that the
low-temperature phase with a few tightly bound defects is destroyed
through unbinding of defect pairs, analogous to the
Kosterlitz-Thouless transition in flat space.  The high-temperature
phase has a finite density of thermally excited, unbound defects.  The
interaction between the defects is screened, with a screening length
of the order of the typical distance of the defects (Debye-H\"uckel
length).  Above the transition temperature, we make a Gaussian
approximation of the Coulomb gas model \ref{cgm} with a
\textit{continuous} defect density $\rho$
\begin{equation}
\fl\frac{H}{T}=\frac{K_{\mathrm{A}}}{2}\int\rmd A\int\rmd A'\;(2\pi\rho-K)_\sigma\;
G(\sigma,\sigma')
\;(2\pi\rho-K)_{\sigma'}+\frac{1}{2x}\int\rmd A\;\rho^2,
\end{equation}
where $x$ is the fugacity of the charges. By setting $\delta H/\delta\rho=0$,
we obtain for the mean charge density
\begin{equation}
\label{dbh}
2\pi\rho=\frac{1}{1-\frac{1}{4\pi^2 K_{\mathrm{A}} x}\Delta}K=K+\frac{1}{4\pi^2 K_{\mathrm{A}} x}
\Delta K+\Or (x^{-2}).
\end{equation}
Although this approximation accurately represents the
Coulomb gas at high temperatures, the use of the Coulomb gas at all to
describe the high-tem\-per\-a\-ture phase is rather suspect. Nevertheless,
we show that the Coulomb gas model yields a density of defects which
agrees remarkably well with the density obtained from \eref{mexican}
in the high-temperature phase on an arbitrary curved
surface.


\section{Charge density in the high-temperature Gaussian approximation}
\label{sec-gauss}

In the remainder of the paper we present the calculation of the defect
density $\rho$ from \eref{mexican} in the disordered state,
where the quartic term $\int\rmd A (u_i u^i)^2$ is irrelevant and can be
neglected.  We expect that similar to the situation in
curved space-time \cite{odintsov},
a term proportional to $\int
dA\;K\;u_iu^i$ is generated under renormalization,
where $K=K(\sigma)$ is the Gaussian
curvature.  Since $K$ has the dimension of $1/\mbox{length}^2$ this
term is as relevant as the gradient term.  Thus, in the high
temperature phase the vector field is distributed according to the
Gaussian weight
\begin{equation}
\label{weight}
\fl P[u]\propto\exp\left(-\frac{1}{2}\int\rmd ^2\sigma\sqrt{g}\left(
D_i u_j D^i u^j+\tau u^i(\sigma) u_i(\sigma) + \eta\;K u^i u_i\right)
\right),
\end{equation}
where $\tau$ is now the mass of the vector field and $\eta$ is the
coupling of $K$ to $u_iu^i$.  In addition, the distribution for $u_i$
has to be equipped with a covariant cutoff procedure, such as the heat
kernel regularization~\cite{camporesi}.  Because the model depends
only on the intrinsic geometry of the manifold, no extrinsic couplings
(such as a term proportional to $C^2 u_i u^i$, where $C$ is the mean curvature
of the surface) can be generated under
renormalization.

The zeros of the field $u_i$ are characterized by the index
$q=\mbox{sign}\det(D_i u_j)=\pm 1$. The index (charge) describes the
local topology of a flow $u_i$ near a zero $u_i(\sigma)=0$.  The
corresponding mean charge density is given by
\begin{equation}
\rho (\sigma) = \left \langle \sum_i q_i\;\delta_{\mathrm{c}}(\sigma,\sigma_i)
\right \rangle,
\end{equation}
where the defects are labeled by the index $i$, located at
coordinates $\sigma_i$, and have charges $q_i$.  The expectation value
is taken with respect to the probability distribution of \eref{weight}.

Transforming from the variable $\sigma$ to the variable $u$ via the
Jacobian, we obtain
\begin{equation}
\label{eq:rho}
\rho=\left<\det(D_iu^j\left(\sigma)\right)
\delta_{\mathrm{c}}\left(u(\sigma)\right)\right>,
\end{equation}
which can be seen easily using a locally Euclidean coordinate system
and linearizing the vector field around the zero
$u_j(x_1,x_2)=x_k\alpha_{kj}$:
\begin{eqnarray}
\fl\int\rmd ^2x\;\det\left(\partial_i(x_k\alpha_{kj})\right)
\delta^2(x_k\alpha_{kj})\nonumber\\
\lo= \int\rmd ^2x\;|\det(\alpha_{ij})|\;\mbox{sign}\det(\alpha_{ij})
\;\delta^2(x_k\alpha_{kj})=\mathrm{sign}\det(\alpha_{ij})=\pm 1.
\end{eqnarray}
To calculate the expectation value (at point $\sigma$) $\rho(\sigma)$
one needs the joint distribution of $u_i(\sigma)$ and $D_i
u_j(\sigma)$ which can be determined since $u_i(\sigma)$ and $D_i
u_j(\sigma)$ are a set of (multicomponent) Gaussian random variables
with correlations $\left<u_i(\sigma) u_j(\sigma)\right>$,
$\left<u_k(\sigma) D_i u_j(\sigma) \right>$, $\left<D_i u_j(\sigma)
D_k u_l(\sigma)\right>$.

\subsection{Density of defects in flat space}
\label{sec-flat}

Before calculating results in curved space, we review Halperin's
calculation for the density of zeros of a Gaussian two-component
order parameter $\bi{u}$ in two dimensional flat
space~\cite{halperin}. \Eref{weight}) becomes
\begin{equation}
P[\bi{u}(\bi{r})] \propto \exp \left\{ -\frac{1}{2}\int\rmd ^2\bi{r}
\left[ (\partial_i u_j)^2 +\tau u^2 \right] \right\}
\label{eq:prob}
\end{equation}
and \eref{eq:rho} becomes
\begin{equation}
\rho (\bi{r}) = \left \langle \delta^2 [\bi{u} (\bi{r})] \det
(\partial_i u_j) \right \rangle.
\end{equation}

This expectation value is completely determined by the probability
distribution $P (\xi_i, \alpha_{ij})$, where $\xi_i = u_i (\bi{r})$
and $\alpha_{ij} = \partial_i u_j (\bi{r})$ via the formula
\begin{equation}
\label{eq:densprob}
\rho (\bi{r}) = \int\rmd ^4 \alpha_{ij} P(\bi{0}, \alpha_{ij})
\det{\alpha_{ij}}.
\end{equation}
Since \eref{eq:prob} is Gaussian, this probability distribution
is just given by
\begin{equation}
\label{eq:matrix}
P (\xi_i, \alpha_{ij}) = \frac{1}{(2\pi)^3} \frac{1}{\left[ \det
M_{ij} \right]^{1/2}} \exp \left\{ - \frac{1}{2} x_i M^{-1}_{ij} x_j
\right\},
\end{equation}
where $\bi{x}$ is a six-component vector given by $\bi{x} = (\xi_1,
\xi_2, \alpha_{11}, \alpha_{12}, \alpha_{21}, \alpha_{22})$, and
$M_{ij}$ is the matrix of correlations $M_{ij} = \langle x_i x_j
\rangle$.  Plugging \eref{eq:matrix} into \eref{eq:densprob} gives
\begin{equation}
\label{eq:flatres}
\rho = \frac{1}{2\pi} \left( \frac{\det \tilde{M}_{ij}}{\det M_{ij}}
\right)^{1/2}\left(\tilde{M}_{14}-\tilde{M}_{23}\right),
\end{equation}
where $\tilde{M}$ is the matrix of correlations $M_{ij} = \langle y_i
y_j \rangle$, and $\bi{y}$ is a four-component vector given by
$\bi{y} = (\alpha_{11}, \alpha_{12}, \alpha_{21}, \alpha_{22})$.

The expectation values necessary to evaluate \eref{eq:flatres}
can be readily determined from \eref{eq:prob}.  The result is
that $\rho(\bi{r}) = 0$, as expected by symmetry: the system is
uniform and charge-neutral.  In order to obtain a nontrivial result to
compare with the Coulomb gas model, we must calculate the correlation
function of the density of charges, defined by
\begin{equation}
\fl C (\bi{r}) = \left \langle \rho(\bi{r}) \rho(\bi{0}) \right \rangle
= \left \langle \delta^2 [\bi{u}(\bi{r})] \det \left[ \partial_i
u_j(\bi{r}) \right] \delta^2 [\bi{u}(\bi{0})] \det \left[
\partial_i u_j(\bi{0}) \right] \right \rangle.
\end{equation}
This can be evaluated by similar methods, and the results match up
very well with the Coulomb gas model~\cite{halperin}.

\subsection{Density of defects in curved space}
\label{sec-curved}

In the case of a general closed membrane, $\rho$ will already be
nontrivial for two reasons.  First, the system is not charge-neutral,
but rather the total charge must be equal to two minus the number of
handles on the surface (e.g. two for a sphere, zero for a
torus, etc.)  Second, unless the surface has a high degree of
symmetry, the charge will not distribute itself uniformly.  Rather,
the charge density will depend upon the local curvature of the
surface.  Thus, for the case we consider in the remainder of the
paper, we can get a meaningful comparison with the Coulomb gas theory
solely from calculating the charge density $\rho$, rather than needing
to calculate the more complicated correlation functions.

The method used is conceptually the same as in \sref{sec-flat},
but more technically complicated due to the curvature of the space.
Therefore, we present it in \ref{sec-details}.  The analogous
result to \eref{eq:flatres} for a \textit{general} Gaussian,
$O(2)$ invariant distribution for the vector field $u_i$ is 
\begin{equation}
\label{dens}
2\pi\rho-K=\epsilon^{ik}\epsilon^{jl}D_k\left(
\frac{\left<(D_i u_j) u_l\right>}{\left<u_m u^m\right>} \right),
\end{equation}
Since the right hand side of \eref{dens} is a total divergence,
we find, after integration over the surface the aforementioned
topological constraint for the total charge of the defects
\begin{equation}
\label{gaussbonnet}
2\pi\int\rmd A\;\rho = \int\rmd A\;K=4\pi(1-\gamma)
\end{equation}
which agrees with the Gauss-Bonnet theorem, as $1-\gamma$ is the genus of
the surface.

To derive $\rho$ from \eref{dens}, we need to calculate
$\left<u_iu_j\right>$ and $\left<(D_iu_j)u_k\right>$.  This can be
done in an expansion with respect to the interaction range $1/\tau$
using the Gaussian weight \eref{weight}.  It is convenient to
decompose the vector field $u_i$ into a sum of a gradient and a curl
$u_i=\partial_i\phi+\epsilon_i{}^j\partial_j\chi$.  This
representation is only valid for (deformed) spheres.  For other
topologies, modes exist which cannot be written as sum of a gradient
and a curl.  For example, in a torus, a vector field that represents a
flow along one of the perimeters cannot be decomposed in this way.

A particularly simple case is given for $\eta=1$ because the
potentials $\phi$ and $\chi$ decouple.
The special role of the $\eta=1$ case can also be understood
within the gauge-field representation,  as shown in \ref{dirac}.
For this case, we derive the density of defects in a high-temperature
expansion.  For high temperatures, the screening length is small compared
to the radius of curvature: the surface appears to be almost flat.
Upon increasing screening length, more details of the geometry become relevant.
We present the details of this high-temperature expansion
in \ref{sec-hightexp}, obtaining as our main result the
average defect density $\rho$:
\begin{equation}
\label{main}2\pi\rho=K+\frac{\Delta K}{12Z\pi\tau}+
\frac{\Delta^2K}{120Z\pi\tau^2}
-\frac{\Delta K^2}{30Z\pi\tau^2}+\Or (\tau^{-3}),
\end{equation}
where $Z^{-1}=2\pi/\log(1/(a^2\tau))$, and $a$ is a short distance
cutoff.

To lowest order in the correlation length $\tau^{-1/2}$, this is
equivalent to the Debye-H\"uckel approximation \eref{dbh} provided
one identifies $\pi K_{\mathrm{A}} x=3Z\tau$.  For larger correlation lengths,
however, deviations show up.  The term $\propto\tau^{-1}$ will be
independent of the coupling $\eta$ for dimensional reasons.  The next
orders, however, depend on $\eta$.  We conjecture, that the expansion
\eref{main} remains valid for arbitrary genus of the surface.
Treating general genus and $\eta$, however, requires the calculation
of moments of the vector field $u_i$ directly, which is much more
complicated and beyond the scope of this work.

It will be difficult to observe the defect density (\ref{main})
experimentally, since $\rho$, which is density of positive defects {\it minus}
the density of negative defects is of the order of $\rho\sim 1/A$ due to
the topological constraint (\ref{gaussbonnet}). On the other hand the density
of positive defects plus the density of negative defects is of the order of
$1/\xi^2\sim\tau$, where $\xi$ is the correlation length, which is small well
above the transition temperature. In the high-temperature phase, therefore,
we have to measure a tiny density difference in the presence of a large
background density. Closer to the transition region the background density
becomes smaller and there might well be a chance to resolve the defects
in thin films
using polarized light. It would be certainly interesting to see wether
our expansion (\ref{main}) or the result (\ref{dbh}) obtained from the
Coulomb gas model allow for the better fit to the experimental data.

\section{Conclusion}

We have derived the average topological charge density of vector
fields with a Gaussian distribution on a curved surface.  We found
that for high temperatures, the zeros behave like (screened) charges
in the presence of a background charge density equal to the Gaussian
curvature.  We demonstrated the validity of the Debye-H\"uckel
approximation of the Coulomb gas model, which, as discussed in
\sref{sec-intro}, is not obvious, since the Coulomb gas model
originates from a low-temperature model of the orientational order and
we are attempting to apply it in a high-temperature regime.

\ack

We have benefited from discussions with D R Nelson and B I Halperin.
GF was supported by the Deutsche Forschungsgemeinschaft through
Grant Fo 259/1 and acknowledges the hospitality of the Condensed
Matter Theory group at Harvard University, where some of this work was
done.  RAL  was supported primarily by the Harvard Materials
Research Science and Engineering Laboratory through Grant
No DMR94-00396, by the National Science Foundation through Grant
No DMR97-14725, and by the Office of Naval Research.

\appendix

\section{Calculation of charge density}
\label{sec-details}

We will calculate the density of defects $\rho=\left<\det(D_iu^j\left(\sigma)
\right)\delta_{\mathrm{c}}\left(u(\sigma)\right)\right>$.  We begin by exact
analogy with the calculation for flat space outlined in
\sref{sec-flat}, by noting that this expectation value is
completely determined by the probability distribution $P (v_i,
A_{ij})$, where $v_i = u_i (\sigma)$ and $A_{ij} = D_i u_j (\sigma)$
via the formula
\begin{equation}
\label{eq:densprob2}
\rho (\sigma) = \int\rmd ^4 A_{ij} P(\bi{0}, A_{ij}) \det{A_i{}^j}.
\end{equation}
Since \eref{weight} is Gaussian, this probability distribution
is just given by
\begin{equation}
\label{eq:matrix2}
P (v_i, A_{ij}) = \frac{1}{(2\pi)^3} \frac{1}{\left[ \det M_{ij}
\right]^{1/2}} \exp \left\{ - \frac{1}{2} x_i \left( M^{-1}
\right)^{ij} x_j \right\},
\end{equation}
where $x_i$ is a six-component vector given by $\left( v_1, v_2,
A_{11}, A_{12}, A_{21}, A_{22} \right)$, and $M_{ij}$ is the matrix of
correlations $M_{ij} = \langle x_i x_j \rangle$.

We can reexpress this as
\begin{equation}
\label{eq:hst}
P (v_i, A_{ij}) = \frac{1}{(2\pi)^6} \int\rmd ^4 \tilde{A}^{ij} \int\rmd ^2
\tilde{v}^{i} \exp \left\{ - \frac{1}{2} \tilde{x}^i M_{ij}
\tilde{x}^j + i \tilde{x}^i x_i \right\},
\end{equation}
where $\tilde{x}^i$ is a six-component vector given by $(\tilde{v}^1,
\tilde{v}^2, \tilde{A}^{11}, \tilde{A}^{12}, \tilde{A}^{21},
\tilde{A}^{22})$.
Performing the integral over the $\tilde{x}^i$
returns us to \eref{eq:matrix2}.
Explicitly we have for the defect density
\begin{eqnarray}
\fl\rho=\frac{1}{(2\pi)^6}\int\rmd ^4\tilde{A}^{ij}d^4A_{ij}
d^2\tilde{v}^id^2v_i
\exp\left(
-\frac{1}{2}\tilde{A}^{ij}\tilde{A}^{kl}\left<(D_iu_j)(D_ku_l)\right>\right)
\nonumber\\
\times\exp\left(
-\tilde{A}^{ij}\tilde{v}^k\left<(D_iu_j)u_k\right>-\frac{1}{2}
\tilde{v}^i\tilde{v}^j\left<u_iu_j\right>+i\tilde{A}^{ij}A_{ij}+i\tilde{v}^i
v_i\right)\nonumber\\
\times\frac{1}{2}\;\epsilon^{ik}\epsilon^{jl}A_{ij}A_{kl}\;\delta_{\mathrm{c}}(v).
\end{eqnarray}
The integration over the $v_i$ yields a factor of $\sqrt{g}$ since
\[\delta_{\mathrm{c}}\left(v(\sigma)\right)=\lim_{\lambda\rightarrow
\infty}
\frac{\lambda}{2\pi}\exp\left(-\frac{\lambda}{2}g^{ij}(\sigma)v_iv_j\right)
=\sqrt{g}(\sigma)\;\delta^2(v).\]
The resulting expression can be
simplified by using the $O(2)$-invariance of the field $u_i$ by noting
that $g_{ij}$ is the only rank two tensor invariant under $O(2)$, and
therefore
\begin{equation}
\left \langle u_i u_j \right \rangle = \frac{1}{2} g_{ij} \left
\langle u_m u^m \right \rangle.
\end{equation}
After integration over the $\tilde{v}^i$ we obtain
\begin{equation}
\label{eq:newprob}
\fl\rho=\frac{1}{2\pi\left<u_mu^m\right>}
\frac{1}{(2\pi)^4}\int\rmd ^4\tilde{A}^{ij}d^4A_{ij}
\exp\left(
-\frac{1}{2}\tilde{A}^{ij}\tilde{A}^{kl}T_{ijkl}
+i\tilde{A}^{ij}A_{ij}\right)
\epsilon^{ik}\epsilon^{jl}A_{ij}A_{kl},
\end{equation}
where
\begin{equation}
T_{ijkl}=\left<(D_iu_j)(D_ku_l)\right>
-2\frac{\left<(D_iu_j)u_m\right>
\left<(D_ku_l)u^m\right>}{\left<u_nu^n\right>}.
\end{equation}
Integrating over the $\tilde{A^{ij}}$ in \eref{eq:newprob},
we see that the  $A_{ij}$ are simply Gaussian variables  
with correlations given by
\begin{equation}
\left \langle A_{ij} A_{kl} \right \rangle = T_{ijkl}.
\end{equation}
Therefore, \eref{eq:newprob} yields
\begin{equation}
\rho=\frac{\epsilon^{ik}\epsilon^{jl}T_{ijkl}}{2\pi\left<u_mu^m\right>}.
\end{equation}

To further simplify this equation, we derive
\begin{eqnarray}
\fl\epsilon^{ik}\epsilon^{jl}D_k\left(\frac{\left<(D_iu_j)u_l\right>}
{\left<u_mu^m\right>}\right)=
\epsilon^{ik}\epsilon^{jl}\frac{\left<(D_kD_iu_j)u_l\right>}
{\left<u_mu^m\right>}
+\epsilon^{ik}\epsilon^{jl}\frac{\left<(D_iu_j)(D_ku_l)\right>}
{\left<u_mu^m\right>}
\nonumber\\
\mbox{}-2\epsilon^{ik}\epsilon^{jl}
\frac{\left<(D_iu_j)u_l\right>\left<(D_ku_m)u^m\right>}
{\left<u_nu^n\right>^2}\nonumber\\
\lo= -K+\epsilon^{ik}\epsilon^{jl}\frac{T_{ijkl}}{\left<u_mu^m\right>}
\end{eqnarray}
using the fact that in two dimensions, the Riemann curvature tensor is
$R_{ijkl} = \epsilon_{ij}\epsilon_{kl}K$, where $K$ is the Gaussian
curvature), and also that due to the $O(2)$-invariance of $u_i$,
\begin{equation}
\left \langle (D_i u_j) u_k \right \rangle = \frac{1}{2} g_{jk} \left
\langle (D_i u_m) u^m \right \rangle + \frac{1}{2} \epsilon_{jk}
\epsilon^{mn} \left \langle (D_i u_m) u_n \right \rangle.
\end{equation}

We therefore arrive at an expression for the mean defect density
\begin{equation}
2\pi\rho-K=\epsilon^{ik}\epsilon^{jl}D_k\left(
\frac{\left<(D_i u_j) u_l\right>}{\left<u_m u^m\right>} \right),
\end{equation}
valid for any Gaussian, $O(2)$-invariant distribution for the vector
fields $u_i$.

\section{high-temperature expansion}
\label{sec-hightexp}

In this Appendix, we derive the density of defects in a
high-temperature expansion for the case $\eta = 1$, as disucssed in
\sref{sec-curved}.  In this case, since $D^i
D_i\partial_j\phi-K\partial_j\phi= \partial_j\Delta\phi$, the
eigenfunctions of the operator $-D_iD^i+K+\tau$ (acting on vector
fields)
\begin{equation}
\left(-D^iD_i+K+\tau\right)u_{j,\alpha}=(\lambda_\alpha+\tau)\;u_{j,\alpha}
\end{equation}
can be written as $u_{i,\alpha}^{(1)}=\partial_i\phi_\alpha$ and
$u_{i,\alpha}^{(2)}=\epsilon_i{}^j\partial_j\phi_\alpha$, where
$\phi_\alpha$ is a normalized eigenfunction of the Laplace-Beltrami
operator
$-\Delta\phi_\alpha=-g^{ij}D_i\partial_j\phi_\alpha=\lambda_\alpha\phi_\alpha$.
Together with the normalization of the $u_\alpha$
\[
\int\rmd A\;u_{j,\alpha}u^{j,\alpha}=
\int\rmd A\;g^{ij}\partial_i\phi_\alpha\partial_j
\phi_\alpha=\lambda_\alpha,
\]
we obtain the propagator for $u_i$
\begin{equation}
\fl\left<u_i(\sigma)u_j(\sigma')\right>=
\left<\partial_i\phi(\sigma)\partial_j\phi(\sigma')\right>
+\epsilon_i{}^k(\sigma)\epsilon_j{}^l(\sigma')
\left<\partial_k\phi(\sigma)\partial_l\phi(\sigma')\right>
\end{equation}
in terms of the {\it scalar} propagator
\begin{equation}
\label{prop}
\left<\phi(\sigma)\phi(\sigma')\right>=
\sum_\alpha{}'\frac{\exp(-a^2\lambda_\alpha)}
{\lambda_\alpha(\lambda_\alpha
+\tau)}\;\phi_\alpha(\sigma)\phi_\alpha(\sigma'),
\end{equation}
where the zero mode $\phi\equiv$const is omitted and $a$ is an
exponential cutoff length that arises from the heat kernel
regularization.  After some algebra we find
\begin{equation}
\label{scalar}
\fl2\pi\rho-K=\epsilon^{ik}\epsilon^{jl}D_k\left(
\frac{\left<(D_i \partial_j\phi)
\partial_l\phi\right>}{\left<\partial_m\phi \partial^m\phi\right>} \right)
=\frac{1}{2}D^k\left(\frac{\partial_k\left<\phi\Delta\phi\right>-
\partial_k\left<\partial_m\phi \partial^m\phi\right>}
{\left<\partial_n\phi \partial^n\phi\right>}\right),
\end{equation}
with $\left<(\Delta\phi)\partial_i\phi\right>=(1/2)\partial_i\left<
(\Delta\phi)\phi\right>$ from \eref{prop}.  Both
$-\left<\phi\Delta\phi\right>$ and
$\left<\partial_m\phi\partial^m\phi\right>$ are logarithmically
divergent for small cutoff lengths
\[
\left.\begin{array}{l}
\left<\partial_m\phi \partial^m\phi\right>\\
-\left<\phi\Delta\phi\right>
\end{array}\right\}=
\frac{1}{4\pi}\log\left(
\frac{1}{a^2\tau}\right)+\mbox{\it finite parts,}
\]
where only the finite parts depend on the position on the manifold.
Therefore the numerator of \eref{scalar} is finite, whereas in
the limit of small cutoff lengths the divergent denominator can be
replaced by its most divergent (spatially constant) part.  Defining
$Z^{-1}=2\pi/\log(1/(a^2\tau))$, we obtain
\begin{equation}
2\pi\rho-K=Z^{-1}\Delta\left(\left<\phi\Delta\phi\right>-\left<
\partial_m\phi\partial^m\phi\right>\right).
\end{equation}

Using
$\left<\partial_m\phi\partial^m\phi\right>=(1/2)\Delta\left<\phi^2
\right>-\left<\phi\Delta\phi\right>$ and $\tau\left<\phi^2\right>=\left<
\phi\Delta\phi\right>+\left<\phi(\tau-\Delta)\phi\right>$, we obtain
\begin{equation}
2\pi\rho-K=Z^{-1}\left(
\Delta\left(2-\frac{1}{2\tau}\Delta\right)\left<\phi\Delta\phi\right>
-\frac{1}{2\tau}\Delta^2\left<\phi(\tau-\Delta)\phi\right>\right).
\end{equation}
The moment 
\[\left<\phi(\tau-\Delta)\phi\right>=
\sum_\alpha{}'\frac{\exp(-a^2\lambda_\alpha)}{\lambda_\alpha}
\;\phi_\alpha(\sigma)^2\]
is the Green's function of the Laplace-Beltrami operator at coinciding
points, which can be obtained from conformal field
theory.
With the definition of the massless propagator
and its short distance behavior
$G(\sigma,\sigma')\sim\Gamma(\sigma)-\log(d(\sigma,\sigma'))/(2\pi)$,
we find for two conformal equivalent metrices
$g_{ij}=\zeta\tilde{g}_{ij}$ that $-\Delta\Gamma+2/A-K/(2\pi)=
(-\tilde{\Delta}\tilde{\Gamma}+2/\tilde{A}-\tilde{K} /(2\pi))/\zeta$
and consequently for a spherical topology
$-\Delta\Gamma=K/(2\pi)-2/A$. $A$ is the area of the surface and $d$
the geodesic distance.
We have
$\Delta^2\left<\phi(\tau-\Delta)\phi\right>=-\Delta K/(2\pi)$.
The moment $\left<\phi\Delta\phi\right>$ has to be calculated in an asymptotic
$1/\tau$ expansion.  With the help of the
references~\cite{camporesi,berger,dewitt} we find for the finite part of
$\left<\phi\Delta\phi\right>$
\begin{equation}
-\left<\phi\Delta\phi\right>=\frac{K}{12\pi\tau}+\frac{K^2+\Delta K}
{60\pi\tau^2}+\Or (\tau^{-3}).
\end{equation}
Finally, we obtain the average defect density $\rho$
\begin{equation}
2\pi\rho=K+\frac{\Delta K}{12Z\pi\tau}+
\frac{\Delta^2K}{120Z\pi\tau^2}
-\frac{\Delta K^2}{30Z\pi\tau^2}+\Or (\tau^{-3}).
\end{equation}

\section{The gauge-field representation}
\label{gauge}
Orientational order is frequently represented as a gauge-field theory
\cite{PLM,Evans}. The vector field $u_i$ is represented in a local orthogonal
base (reference frame) 
$v_i$ by a complex function $\psi$ through 
$u_i=v_i\;\mbox{Re}(\psi) + \epsilon_i{}^jv_j\;\mbox{Im}(\psi)$, where
$v_iv^i=1$ and $\epsilon$ is the antisymmetric unit tensor.
Plugging this into \eref{mexican}, we obtain
\begin{equation}
\label{eq:gauge}
\fl\frac{H}{T}= \int\rmd ^2\sigma\sqrt{g}\left(g^{ij}
\left( \partial_i\psi^* + i \Omega_i \psi^* \right) \left(
\partial_j\psi - i \Omega_j\psi \right)+\tau|\psi|^2+c|\psi|^4\right)
\end{equation}
with the vector potential $\Omega_i=\epsilon^{jk}v_jD_iv_k$, resulting from the
fact that the reference frame $v_i$ is changing from point to point.
In this representation the Gaussian curvature $K$ plays the role of a
perpendicular magnetic field $\epsilon^{ij}D_i\Omega_j=K$.  Since any
unit vector field $v_i$ must have two points on a sphere where it is
singular, $\Omega_i$ will be singular at two points as well, even if the
underlying surface is not.  We therefore use \eref{mexican}
rather than \eref{eq:gauge} in this paper as a base for
calculation.  The disadvantage of \eref{mexican} is that it is
more difficult to generalize to an $n$-fold symmetry, as is done by
Park, \textit{et\ al\ }~\cite{park1} and Evans~\cite{Evans}.

\section{The $\eta=1$ case}
\label{dirac}

\Eref{eq:gauge}) is covariant and gauge invariant,
{\it i.e.} invariant against changes of the local frame $v_i$.
For convenience, we choose a conformally flat coordinate system with a
metric tensor $g_{ij}=\zeta(x,y)\delta_{ij}$, where the coordinates
are  $\sigma^1\equiv x, \sigma^2\equiv y$ and $\zeta(x,y)$ encodes
the (intrinsic) geometry of the surface.
A proof that such a coordinate system exists, as
well as a review of its properties, can be found in \cite{nels}.
(Note, however, that for closed surfaces other than tori, $\zeta$ will
have singular points.) Furthermore, we choose a particular reference frame 
$v_x=\sqrt{\zeta},\;v_y=0$. Then $\sqrt{g}g^{ij}=\delta_{ij}$, $\Omega_x=
\partial_y\omega$, $\Omega_y=-\partial_x\omega$,
$K=(1/\zeta)(\partial_x^2+\partial_y^2)\omega$ where $\omega=-(1/2)\log\zeta$.
The Gaussian weight \eref{weight} becomes
$P[\psi]=\exp(-1/2\;\int\rmd ^2x\;\psi^*{\cal H}\psi)$
with the Hamilton operator
\begin{equation}
{\cal H}=-(\partial_x-i\partial_y\omega)^2-(\partial_y+i\partial_x\omega)^2+
\eta\nabla^2\omega+\zeta\tau.
\end{equation}
For $\eta=1$ and a vanishing mass $\tau=0$ we can express the Hamilton
operator in terms of the square of
a Dirac-type operator ($\hat{\sigma}_{x,y,z}$ are the Pauli-matrices)
\begin{eqnarray}
\fl{\cal H}=
-\left[\left(\partial_x-i\partial_y\omega\right)\hat{\sigma}_x+
\left(\partial_y+i\partial_x\omega\right)\hat{\sigma}_y\right]^2
\nonumber\\
\lo= -(\partial_x-i\partial_y\omega)^2-(\partial_y+i\partial_x\omega)^2
+\hat{\sigma}_z\nabla^2\omega
\end{eqnarray}
in the $\sigma_z=+1$ sector. The latter operator is the Hamilonian of the
2D-Pauli equation for spin$-1/2$ particles with the (dimensionless)
magnetic moment $g=2$ (electrons!) in a magnetic field $B_z=\nabla^2\omega$.  
Thus the $\eta=1$ case that we focus on is closely related to both the Dirac
equation and the Pauli equation in two dimensions. The Dirac
equation and its discretized counterpart are commonly used to describe the
properties of electrons confined to a plane in a quenched, perpendicular 
magnetic field \cite{ludwig}.

\Figures

\end{document}